\documentclass[letterpaper, 10pt,reqno]{amsart}

\usepackage[usenames,dvipsnames]{xcolor}

\usepackage[portrait,margin=3cm]{geometry}
\usepackage{mathrsfs}

\usepackage{amssymb,amsthm,amsmath,amsfonts,amsbsy,latexsym}

\usepackage{color}
\usepackage{graphicx}
\usepackage{bbm}
\usepackage{comment}
\usepackage{a4wide, hyperref,graphicx, bbm, eurosym, listings, float, graphicx, amsfonts, comment,lscape, epstopdf, units, algorithmic, algorithm, textcomp, amsthm, fancyhdr, url, color, soul, comment, enumerate, color,tikz}
\usepackage{enumitem}

\usepackage{mathtools}
\mathtoolsset{showonlyrefs}

\usepackage{subcaption}

\definecolor{MyDarkblue}{rgb}{0,0.08,0.50}
\definecolor{Brickred}{rgb}{0.65,0.08,0}

\hypersetup{
	colorlinks=true,       
	linkcolor=MyDarkblue,          
	citecolor=Brickred,        
	filecolor=red,      
	urlcolor=cyan           
}

\newtheorem*{theorem*}{Theorem}

\newcommand{\N}{\mathbb{N}}
\newcommand{\Z}{\mathbb{Z}}

\renewcommand{\emptyset}{\varnothing}

\newcommand*{\wt}{\widetilde}

\newcommand*{\be}{\begin{equation}}
\newcommand*{\ee}{\end{equation}}
\newcommand*{\ba}{\begin{aligned}}
	\newcommand*{\ea}{\end{aligned}}
\newcommand*{\barr}{\begin{array}{c}}
	\newcommand*{\earr}{\end{array}}

\def\namedlabel#1#2{\begingroup
	#2%
	\def\@currentlabel{#2}%
	\phantomsection\label{#1}\endgroup
}

\newcommand{\bes}{\begin{equation*}}
\newcommand{\ees}{\end{equation*}}

\renewcommand{\N}{\mathbb{N}}

\numberwithin{equation}{section}

\begin{document}
	\title{Mapping NP-hard and NP-complete optimisation problems to Quadratic Unconstrained Binary Optimisation problems.}
	
	\date{\today}
	
	\author[Lodewijks]{Bas Lodewijks}
	\address{Department of Mathematical Sciences,
		University of Bath,
		Claverton Down,
		Bath,
		BA2 7AY,
		United Kingdom.}
	\email{b.lodewijks@bath.ac.uk}
	\pagestyle{plain} 
	\maketitle 

\begin{abstract} 
  We discuss several mappings from well-known NP-hard problems to Quadratic Unconstrained Binary Optimisation problems which are treated incorrectly by Lucas in \cite{Luc14}. We provide counterexamples and correct the mappings. We also extend the body of QUBO formulations of NP-complete and NP-hard optimisation problems by discussing additional problems.
\end{abstract} 

\section{Introduction}
In recent years there has been an increasing interest in the use of Adiabatic Quantum Optimization (AQO) to solve optimisation problems, mostly focussing on NP-complete and NP-hard problems \cite{Far01}. As it is agreed upon by most that solving such optimisation problems requires an exponentially large runtime using classical computing techniques, other approaches that can improve on the existing algorithms are highly sought after, especially due to many applications of such problems.\\

Quantum Annealing (QA) devices which implement the AQO technique, which is based on the adiabatic principle of quantum mechanics \cite{FeoRes89}, have been developed and some experimental results seem to provide evidence that shows a speed-up compared to the existing techniques \cite{HeRoIsa15,RoWangJob14}, though there are theoretical arguments that show how an exponential runtime is still to be expected as well \cite{Alt10,DickAmin11}.\\

In order to use AQO, a formulation of an optimisation problem in terms of a Quadratic Unconstrained Binary Optimisation (QUBO) problem is required, as well as a proper embedding of this QUBO in the hardware of the QA devices. A QUBO problem deals with minimising a polynomial function of binary variables, with degree at most two \cite{BorHam02}. That is, 
\be \label{eq:qubo}
\min_{x\in\{0,1\}^N}\sum_{i=1}^N h_i x_i+\sum_{1\leq i<j\leq N} J_{ij}x_ix_j,
\ee 
for some $N\in\N$ and coefficients $(h_i)_{1\leq i\leq N},(J_{ij})_{1\leq i<j\leq N}$. It is possible for most optimisation problems to be caught in a Polynomial Unconstrained Binary Optimization (PUBO) problem, but a QUBO formulation requires more effort and is more challenging. In \cite{Luc14}, Lucas provides an overview of how to map many well-known NP-complete and NP-hard problems into a QUBO formulation. Note that the form of \eqref{eq:qubo}, when applying the straightforward mapping $x_i\mapsto 2x_i-1$, yields the minimisation of a Hamiltonian of an Ising model, which is an NP-complete problem itself \cite{Bar82}.\\

When a QUBO formulation is obtained, the next challenge is to embed the formulation onto the hardware of QA devices. Due to the specific form of the Chimera graphs used in QA devices and the technical limitations still present, regarding the number of qubits in QA devices, it can be challenging to properly embed particular problems onto the Chimera graphs used in these devices \cite{Luc19,RiefVent15,VentMad15}. \\

\textbf{Our contribution}\\
In this paper we deal with the transformation of certain NP-complete and NP-hard problems into QUBO problems. In the overview presented in \cite{Luc14}, some of the transformations are either incorrect or can be simplified. That is, when solving the original NP-complete or NP-hard problem and solving its QUBO equivalence as proposed in \cite{Luc14}, the solutions do not agree. Hence, the QUBO does not solve the same optimisation problem and therefore is obtained via an incorrect mapping of the original problem.

We point out which transformations are incorrect in Section \ref{sec:correction} and provide examples where things go wrong. We then correct the transformations such that the mapping from the particular NP-hard problems into QUBO problems always yields the correct solution. In Section \ref{sec:map}, we provide additional transformations from various other NP-complete and NP-hard problems to QUBOs.

\section{Correcting QUBO mappings}\label{sec:correction}

In this section we present the NP-hard problems discussed in \cite{Luc14} for which an incorrect mapping to a QUBO problem has been constructed. We use similar terminology, in the sense that we call the cost function that is to be minimised a Hamiltonian, as we can interpret QUBOs as minimising an Ising model Hamiltonian, and we denote by a configuration a particular realisation of the binary variables. In all problems, whenever we provide Hamiltonians in terms of functions $H_A,H_B,H_C$, the Hamiltonian we construct is always the sum of these components, i.e.\ $H=H_A+H_B+H_C$.

\subsection{Clique problem}
The clique problem deals with two optimization problems: the NP-complete problem which answers the question `Given a graph $G=(V,E)$, does a clique of size $K$ exist in the graph?' and the NP-hard problems which attempts to find the largest clique in the graph $G$. The former is dealt with correctly in \cite{Luc14}, but the latter is not. Lucas introduces the binary variables $x_v,v\in V,y_i,i\in\{2,\ldots,\Delta\}$, where $\Delta:=\max_{v\in V}d_v$, with $d_v$ the degree of $v$. Here, $x_v$ equals $1$ if $v$ is part of the largest clique, and $y_i$ equals $1$ if the largest clique is of size $i$. Then, the Hamiltonian is constructed as follows.
\be \ba\label{eq:incclique}
H_A &= A\Big(1-\sum_{i=2}^\Delta y_i\Big)^2 + A\Big(\sum_{i=2}^{\Delta} iy_i - \sum_{v\in V} x_v\Big)^2,\\
H_B &= B\Big[\frac{1}{2}\Big(\sum_{i=2}^{\Delta} i y_i\Big)\Big(\sum_{i=2}^{\Delta} iy_i-1\Big) - \sum_{(u,v)\in E} x_u x_v\Big],\\
H_C &= -C\sum_v x_v,
\ea \ee  
and Hamiltonian is $H=H_A+H_B+H_C$. The terms in $H_A$ ensure that the largest clique has a unique size and that the size encoded by the $y_i$ variables matches the number of vertices part of the largest clique, as encoded by the $x_v$ variables. Then, $H_B$ penalises configurations in which the number of edges connecting the vertices $v\in V$ such that $x_v=1$ is too low. Finally, $H_C$ ensures the largest clique is found. According to Lucas, the constants $A,B,C$ should satisfy the following constraints:
\be\label{eq:cliqueconst}
A>\Delta B, \qquad C<A-\Delta B, \qquad A,B,C>0.
\ee 
There are two mistakes in this formulation. First, the variables $y_i$ should be defined for $i\in\{2,\ldots,\Delta+1\}$, that is, the largest clique that can exist in a graph is of size $\Delta+1$, not $\Delta$. This extra variable should be included accordingly in all sums in $H_A,H_B$ in \eqref{eq:incclique}, with $i$ going up to $\Delta+1$. Thus, the correct version is
\be \ba\label{eq:clique}
H_A &= A\Big(1-\sum_{i=2}^{\Delta+1} y_i\Big)^2 + A\Big(\sum_{i=2}^{\Delta+1} iy_i - \sum_{v\in V} x_v\Big)^2,\\
H_B &= B\Big[\frac{1}{2}\Big(\sum_{i=2}^{\Delta+1} i y_i\Big)\Big(\sum_{i=2}^{\Delta+1} iy_i-1\Big) - \sum_{(u,v)\in E} x_u x_v\Big],\\
H_C &= -C\sum_v x_v.
\ea \ee  
Then, the constraints on the constants $A,B,C$ are not strict enough. It is possible to choose the constants such that $B\leq C$ (Lucas own suggestions for the choice of the constants is $B=C$), which can lead to an optimal solution for the QUBO, which is not an optimal solution, or even a clique for that matter, in the clique problem. The following example shows how this can happen. 

Let $G=(V,E)$, with $V=\{1,2,3,4\}$, $E=\{(1,2),(1,3),(1,4),(2,3),(3,4)\}$ be a graph and set $B>0,A=5B,C\in[B,2B)$. When inspecting the graph, the optimal solution should be a clique of size $3$, which would translate into a value $H=-3B$. However, when we set $y_4=1$ and $x_i=1$ for all $i\in V$, it follows that $H_A=0,H_B=B$. But, since we have $4$ vertices in the `clique', $H_C=-4C$, which yields an optimal value of $H=B-4C\leq -3B$. So, when $B\leq C$, there exists an optimal solution to the QUBO problem which is not a solution to the clique problem, let alone the optimal solution.

It becomes clear that, when $B\leq C$, an optimal solution for the QUBO can yield a selection of vertices that do not form a clique, as the Hamiltonian enforces a penalty due to some edges that are missing, which is made up for by allowing more vertices in the clique. This can be countered by increasing the penalty on missing edges, so choosing $B>C$. Therefore, the constants' constraints should be 
\be 
A>\Delta B, \qquad C<\min(A-\Delta B,B), \qquad A,B,C>0.
\ee 

\subsection{Log reduction of auxiliary variables}

For certain problems, it is possible to reduce the number of variables required in the QUBO formulation. In the clique problem, one can encode the clique size, as used in $H_A,H_B$ in \eqref{eq:clique}, in a different way. This method uses the idea of logarithmic encoding \cite{Luc14}. Given $N$ variables $y_i,i\in\{0,\ldots,N\}$, we introduce variables $\wt y_i,i\in\{1,\ldots,M\}$, where $M:=\lfloor \log N\rfloor$. Now, the following two expressions encode the same values:
\be \label{eq:log}
\sum_{i=1}^N iy_i,\qquad \sum_{i=0}^{N-1}2^i \wt y_i +(N+1-2^M)\wt y_N.
\ee   
Note that, when using the log encoding, the first term of $H_A$ in \eqref{eq:clique} can be omitted, as it is no longer required to only allow for one $\wt y_i$ to equal $1$.

In order to use this approach for the clique problem, a small adjustment is necessary, as the first term in \eqref{eq:log} appears in \eqref{eq:clique}, but with the index ranging from $i=2$ to $\Delta+1$. So, we set $M=\lfloor \log (\Delta+1)\rfloor$ and use
\be 
1+\sum_{i=0}^{\Delta}2^i \wt y_i +(\Delta+2-2^M)\wt y_N.
\ee 
This term maps the $\wt y_i$ to the values $2,\ldots,\Delta+2$, whereas we originally could only use the values $2,\ldots,\Delta+1$. However, setting all $\wt y_i$ to equal $1$, which leads to the value $\Delta+2$, yields a Hamiltonian value of at least $B-C>0$, which therefore never is an optimal solution.

We note that in all other problems discussed in \cite{Luc14} where a log reduction can be applied to a term as the first term in $\eqref{eq:log}$, the original approach is valid.

\subsection{Graph colouring problem}
In the graph colouring problem, given a graph $G=(V,E)$ and a set of $n$ different colours, the aim is to find a colouring of the vertices of the graph $G$ such that every pair of vertices that are connected by an edge have a different colour. Lucas presents the Hamiltonian
\be \label{eq:graphcol}
H = A\sum_{v\in V} \Big(1-\sum_{i=1}^n x_{v,i}\Big)^2 + A\sum_{(u,v)\in E} \sum_{i=1}^n x_{u,i}x_{v,i},
\ee 
where $x_{v,i}$ equals $1$ if vertex $v$ has colour $i$, $v \in V,i\in\{1,\ldots,n\}$. The first term ensures that each vertex has a unique colour, and the second term penalises colourings such that connected vertices have the same colour. There is a small mistake, as the two terms should have a different constant weight, rather than both having weight $A$. Now, it is possible to allow for a penalty $A$ in the first term by assigning no colour to a vertex, which can then lead to at most a $\Delta A$ reduction of the second term, where $\Delta$ is the maximum degree in the graph. This might, for example, happen when a graph $G$ is not $n$-colourable and there exists a vertex $v$ that has at least $n$ neighbours such that for every colour, there is neighbour with that colour. 

As an example, set $n=2$ and consider the complete graph of size $5$. The optimal solution should be one where $3$ vertices have colour $1$, say vertices $1,3,5$ and two vertices have colour $2$, say vertices $2,4$. The minimum of $H$, however, is realised when, for example, vertices $1,3$ have colour $1$ and vertices $2,4$ have colour $2$, and vertex $5$ has no colour. The optimal colouring has a value $H=4A$, whereas when assigning no colour to vertex $5$, that is, $x_{5,1}=x_{5,2}=0$, $H=3A$ can be achieved. 

In order to overcome this problem, the second term should be weighted with a constant $B>0$, where $A>\Delta B$. So, the Hamiltonian then becomes
\be 
H = A\sum_{v\in V} \Big(1-\sum_{i=1}^n x_{v,i}\Big)^2 + B\sum_{(u,v)\in E} \sum_{i=1}^n x_{u,i}x_{v,i}.
\ee
In the specific case of $n=2$, we provide a different formulation, as it is now possible to encode the colour of a vertex within a single binary variable. Let $x_v,v\in V$ denote the colour variables, where $x_v$ equals $0,1$ if $v$ has colour $0,1$, respectively. Then, we find
\be 
H=\sum_{(u,v)\in E}1-(x_u+x_v-2x_ux_v).
\ee 
It is clear that a penalty is invoked if and only if the colours of vertices $u,v$ that are connected by an edge do not match. 

\subsection{Tree problems}
\subsubsection{Minimal Spanning Trees with a maximal degree constraint}\label{sec:mst}
Given a connected, undirected graph $G=(V,E)$ with positive edge-weights $\{c_e\}_{e\in E}$, the Minimal Spanning Tree (MST) is a tree $T=(V,E_T)$, such that $E_T\subseteq E$ and $\sum_{e\in E_T}c_e$ is minimised. Finding an MST is not hard (it is in P rather than NP), but when an extra constraint relating to the largest degree $\Delta$ that is allowed in the MST is included, the problem becomes NP-hard.

Lucas aims to find the MST with the maximal degree constraint by tracking the depth of each vertex in the MST. The depth of a vertex in a tree is its graph distance to the root. So, let $x_{v,i},v\in V,i\in\{0,\ldots,\lfloor N/2\rfloor\}$ be a variable which equals $1$ when vertex $v$ has depth $i$ in the MST. Then, also define variables $x_{uv,i},x_{vu,i},(u,v)\in E,i\in\{1,\ldots,\lfloor N/2\rfloor\}$, where  $x_{uv,i}$ (resp.\ $x_{vu,i}$) equals $1$ if edge $(u,v)$ is part of the MST and $x_{u,i-1},x_{v,i}=1$ (resp.\ $x_{v,i-1},x_{u,i}=1$), i.e.\ it connects a vertex of depth $i-1$ to a vertex of depth $i$. Furthermore, $y_{uv},(u,v)\in E$, is a variable which equals $1$ if $(u,v)$ is part of the MST. Finally, Lucas introduces variables $z_{v,j},v\in V,j\in\{1,\ldots,\Delta\}$, such that $z_{v,j}$ equals $1$ if vertex $v$ has exactly $j$ neighbours in the MST. Then, the Hamiltonian presented by Lucas is of the form
\be \ba\label{eq:mst}
H_A={}& A\Big(1-\sum_{v\in V} x_{v,0}\Big)^2 + A\sum_{v\in V} \Big(1-\sum_{i=0}^{\lfloor N/2\rfloor} x_{v,i}\Big)^2 \\
&+ A\sum_{(u,v)\in E} \Big(y_{uv}-\sum_{i=1}^{\lfloor N/2\rfloor} (x_{uv,i}+x_{vu,i})\Big)^2 + A\sum_{v\in V} \sum_{i=1}^{\lfloor N/2\rfloor}\Big(x_{v,i}-\sum_{u:(u,v)\in E} x_{uv,i}\Big)^2 
\\&+ A \sum_{v\in V} \Big(\sum_{j=1}^\Delta jz_{v,j}-\sum_{u:(u,v)\in E}\sum_{i=1}^{\lfloor N/2\rfloor} (x_{uv,i}+x_{vu,i})\Big)^2\\
&+A\sum_{(u,v)\in E} \sum_{i=1}^{\lfloor N/2\rfloor} x_{uv,i}(2-x_{u,i-1}-x_{v,i})+x_{vu,i}(2-x_{v,i-1}-x_{u,i}),\\
H_B ={}&B\sum_{(u,v)\in E}\sum_{i=1}^{\lfloor N/2\rfloor } c_{uv}(x_{uv,i}+x_{vu,i}),
\ea \ee
with $c_{uv}$ the edge-weight of edge $(u,v)$ and  $A>\max_{e\in E}c_e B,B>0$. The many terms in the Hamiltonian all have their own purpose. The first and second term ensure the MST has a unique root and each vertex has a unique depth, respectively. The third term ensures that an edge has a unique depth if and only if it is part of the MST and no depth otherwise. The fourth term enforces the tree structure: it makes sure every vertex in the tree has a unique parent vertex, i.e.\ every vertex, except the root, is connected to exactly one vertex that is closer to the root. The fifth term deals with the maximal degree constraint, so that no vertex has more connections within the MST than the maximal degree constraint allows, namely $\Delta$ many. The last term of $H_A$ forces vertex and edges to have matching depths: for an edge $(u,v)$, it should be included within the tree with depth $i$ only if vertex $v$ has depth $i-1$ and vertex $u$ of depth $i$ or vice versa. Finally, $H_B$ minimises the objective, the total cost of the tree. For the constants, $A>B\max_{e\in E}c_e ,B>0,$ should be satisfied.\\

\begin{figure}[h]
	\centering
	\begin{subfigure}{0.5\textwidth}
		\centering
		\includegraphics[width=0.6\textwidth]{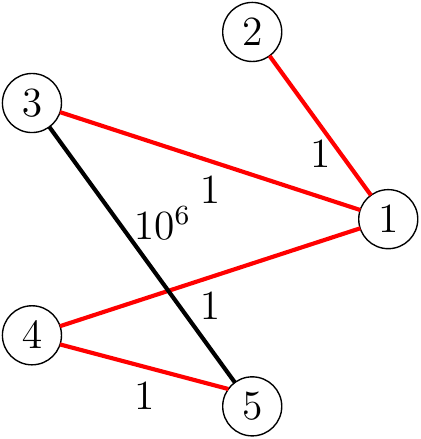}
		\caption{The optimal solution to the incorrect QUBO.}\label{fig:mst}
	\end{subfigure}%
	\begin{subfigure}{0.5\textwidth}
		\centering
		\includegraphics[width=0.6\textwidth]{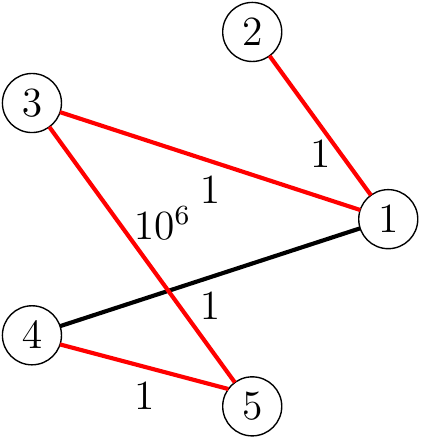}
		\caption{The optimal solution to the correct QUBO.}\label{fig:mst_correct}
	\end{subfigure}
	\caption{Left: the Minimum Spanning Tree in red, which is the optimal solution to the QUBO problem in \eqref{eq:mst}, though an incorrect solution to the MST with maximum degree constraint problem. Right: the Minimum Spanning Tree in red, which is the optimal solution to the QUBO problem in \eqref{eq:mstcorrect} as well as to the MST with maximum degree constraint problem.}
\end{figure}

There are two things that can be improved with respect to this formulation. First, and most importantly, this Hamiltonian does not enforce the maximal degree constraint properly. The term in the third line of $H_A$ is meant to penalise every vertex that has a degree greater than $\Delta$. However, for every vertex $v\in V$, it is possible to allow multiple $z_{v,j}$ to equal $1$, in which case degrees larger than $\Delta$ can be obtained without a penalty. This is clarified in the following example: let $G=(V,E)$, with $V=\{1,2,3,4,5\}$, $E=\{(1,2),(1,3),(1,4),(3,5),(4,5)\}$, let $(c_{1,2},c_{1,3},c_{1,4},c_{3,5},c_{4,5})=(1,1,1,10^6,1)$ and let $\Delta=2$. By the maximum degree constraint, we should only find optimal solutions that are paths, but as becomes clear in Figure \ref{fig:mst}, the high cost of the edge $(3,5)$ makes it more favourable to ignore the maximum degree constraint, which is possible by the above, and to not include the edge with the $10^6$ cost (note that any path covering all vertices must go through the edge $(3,5)$). The solution is to include the term 
\be 
A\sum_{v\in V} \Big(1-\sum_{j=1}^\Delta z_{v,j}\Big)^2
\ee 
to $H_A$, which enforces the degrees to indeed be at most $\Delta$, as for every $v\in V$, exactly one $z_{v,j}$ equals $1$.

The other issue is that the Hamiltonian can be simplified. The first term on the second line of \eqref{eq:mst} is redundant, as the variables $y_{uv}$ play no role elsewhere. They only need to match the sum included in the term. A penalty is invoked only when the sum has a value greater than $1$, but in that case the other terms in $H_A$ already invoke penalties as well. Therefore, this term can be omitted. Thus, the correct Hamiltonian is
\be\ba\label{eq:mstcorrect}
H_A={}& A\Big(1-\sum_{v\in V} x_{v,0}\Big)^2 + A\sum_{v\in V} \Big(1-\sum_{i=0}^{\lfloor N/2\rfloor} x_{v,i}\Big)^2 \\
&+ A\sum_{v\in V} \sum_{i=1}^{\lfloor N/2\rfloor}\Big(x_{v,i}-\sum_{u:(u,v)\in E} x_{uv,i}\Big)^2+A\sum_{v\in V} \Big(1-\sum_{j=1}^\Delta z_{v,j}\Big)^2 
\\&+ A \sum_{v\in V} \Big(\sum_{j=1}^\Delta jz_{v,j}-\sum_{u:(u,v)\in E}\sum_{i=1}^{\lfloor N/2\rfloor} (x_{uv,i}+x_{vu,i})\Big)^2\\
&+A\sum_{(u,v)\in E} \sum_{i=1}^{\lfloor N/2\rfloor} x_{uv,i}(2-x_{u,i-1}-x_{v,i})+x_{vu,i}(2-x_{v,i-1}-x_{u,i}),\\
H_B ={}&B\sum_{(u,v)\in E}\sum_{i=1}^{\lfloor N/2\rfloor } c_{uv}(x_{uv,i}+x_{vu,i}),
\ea\ee 
with $A>\max_{e\in E}c_e B,B>0$. This Hamiltonian does realise the correct solution, as can be seen in Figure \ref{fig:mst_correct}.\\

\subsubsection{Steiner trees}\label{sec:stein}

In the Steiner tree problem the aim is, given an undirected graph $G=(V,E)$ and a set of vertices $U\subseteq V$, to find a spanning tree that contains (at least) all vertices in $U$. When $U=V$, this problem comes down to finding an MST of the graph $G$, when $U$ contains exactly two vertices $u,v$, the problem reduces to finding the shortest path from $u$ to $v$. Though both of these problems, finding a MST and finding the shortest path, are in the complexity class P, the general problem of finding a Steiner tree, which interpolates between these two problems, is in NP.\\

Lucas states that the Steiner tree problem is very similar to the problem in Section \ref{sec:mst}, only now we no longer need to worry about maximal degrees, but rather which vertices to include in the spanning tree. To this end, Lucas uses the same variables as for the MST with maximal degree constraint problem in Section \ref{sec:mst}, and introduces new variables $y_v,v\in V\backslash U$, where $y_v=1$ if $v$ is included in the MST. Then, Lucas provides the following Hamiltonian:
\be \ba \label{eq:stein}
H_A={}& A\Big(1-\sum_{v\in V} x_{v,0}\Big)^2   + A\sum_{v\in U} \Big(1-\sum_{i=0}^{\lfloor N/2\rfloor} x_{v,i}\Big)^2 + A\sum_{v\notin U} \Big(y_v-\sum_{i=0}^{\lfloor N/2\rfloor} x_{v,i}\Big)^2 \\
&+ A\sum_{v\in V} \sum_{i=1}^{\lfloor N/2\rfloor}\Big(x_{v,i}-\sum_{(u,v)\in E} x_{uv,i}\Big)^2 \\
&+ A\sum_{(u,v)\in E} \sum_{i=1}^{\lfloor N/2\rfloor} x_{uv,i}(2-x_{u,i-1}-x_{v,i})+x_{vu,i}(2-x_{v,i-1}-x_{u,i})\\
&+ A\sum_{(u,v)\in E} \Big(y_{uv}-\sum_{i=1}^{\lfloor N/2\rfloor} (x_{uv,i}+x_{vu,i})\Big)^2.
\ea\ee
As in the MST with maximal degree constraint problem, the last term is not necessary and therefore can be omitted.

\subsubsection{Undirected feedback set}\label{sec:undfeedvert}

Given an undirected graph $G=(V,E)$, the undirected feedback set problem finds the smallest set $F\subseteq V$ such that the induced graph on $V\backslash F$ is acyclic. Again, this problem is related to the MST with maximal degree constraint problem and the Steiner tree problem as in Sections \ref{sec:mst}, \ref{sec:stein}, respectively, and similar variables as in Section \ref{sec:stein} are used, where now $y_v,v\in V$ equals $1$ if $v$ is part of the feedback set and $y_{uv},y_{vu},(u,v)\in E$ equal $1$ if $v\in F$ and $u\in F$, respectively. That is, the edge $(u,v)$, $(v,u)$, respectively, points `towards' a vertex in the feedback set. Note that due to the fact that the graph is undirected, the orientation $(u,v),(v,u)$ is not important. The Hamiltonian, according to Lucas, then should be
\be\ba\label{eq:undfeedvert}
H_A ={}& A\sum_{v\in V} \Big(1-y_v-\sum_{i=0}^{\lfloor N/2\rfloor} x_{v,i}\Big)^2 + A\sum_{(u,v)\in E} \Big(1-\sum_{i=1}^{\lfloor N/2\rfloor} (x_{uv,i} + x_{vu,i}+y_{uv}+y_{vu})\Big)^2 \\
&+ A \sum_{(u,v) \in E} (y_{uv}-y_{v})^2 + A\sum_{v\in V}\sum_{i=1}^{\lfloor N/2\rfloor}\Big(x_{v,i} - \sum_{u:(u,v)\in E}x_{uv,i}\Big)^2\\
&+ A\sum_{(u,v)\in E} \sum_{i=1}^{\lfloor N/2\rfloor} x_{uv,i}(2-x_{u,i-1}-x_{v,i})+x_{vu,i}(2-x_{v,i-1}-x_{u,i}),\\
H_B={}& B\sum_{v \in V} y_v,
\ea\ee 
where $A>B>0$. The Hamiltonian aims to find a `spanning forest', that is, a single tree is no longer required, and excludes the vertices and its adjacent edges which would make the graph $G$ cyclic. The first terms denotes that, in case $v\in F$ ($y_v=1$), it should not have a depth in the spanning tree and a unique depth otherwise. Likewise, for the second term, if $y_{uv}$ and/or $y_{vu}$ equals $1$, the edge has an endpoint in the feedback set and therefore should not be included in the `spanning forest'. This implies that the edge should not have a depth in the tree. The third term aims to ensure that, if a vertex is part of the feedback set, its adjacent edges should not be included in the `forest'. Finally, the last two terms are similar to the terms in \eqref{eq:mstcorrect}, \eqref{eq:stein}, and deal with the correct depth structure of the tree.\\

There are a few mistakes in this Hamiltonian. First, the inner summation of the second term of $H_A$ should not include the term $y_{uv}+y_{vu}$, this should be outside of the summation. Second, note that if we swap the orientation of all edges, i.e.\ $(u,v)$ becomes $(v,u)$, we still have the same graph, as it is undirected. Therefore, the Hamiltonian should be invariant to this change, but this is not the case. Hence, for any connected cyclic graph $G=(V,E)$, there exist optimal solutions to \eqref{eq:undfeedvert} that satisfy $F=\emptyset$ and $H=0$, which does not solve the original problem. 

This is possible due to the second (with the correction mentioned above) and third term in \eqref{eq:undfeedvert}. Namely, we can set, for all $(u,v)\in E,\ y_{uv}=y_v=0$. Then, the Hamiltonian is almost equal to the one of the MST problem in \eqref{eq:mstcorrect}, but without the maximum degree constraint terms (the fourth and fifth term of $H_A$ in \eqref{eq:mstcorrect}). Then, for any spanning tree of $G$, set $y_{vu}=1$ for any edge $(u,v)\in E$ that is not part of the spanning tree. It follows that $H=0$ for such configurations. So, for any connected graph $G$ there is an optimal solution $F=\emptyset,H=0$, including cyclic graphs, which does not match the original problem.

The way to resolve this issue is by including a term
\be \label{eq:symterm}
A\sum_{(u,v)\in E} (y_{vu}-y_u)^2,
\ee 
which mirrors the third term in \eqref{eq:undfeedvert}. When including this term, if $G$ is cyclic, we can no longer set all $y_v$ to $0$ without being penalised by this term or the second term in \eqref{eq:undfeedvert}, as this term restores the symmetry in the Hamiltonian.\\

\begin{figure}[h]
	\centering
	\begin{subfigure}{0.5\textwidth}
		\centering
		\includegraphics[width=0.6\textwidth]{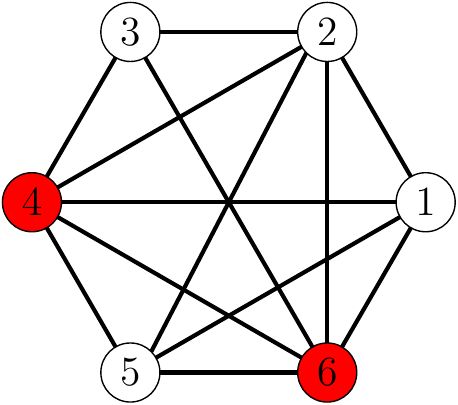}
		\caption{The optimal solution to the incorrect QUBO, when it includes the term in \eqref{eq:symterm}.}\label{fig:undfeedvertex_wrong}
	\end{subfigure}%
	\begin{subfigure}{0.5\textwidth}
		\centering
		\vspace{-10pt}
		\includegraphics[width=0.6\textwidth]{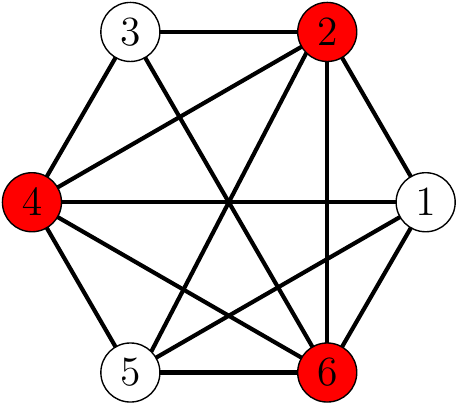}
		\caption{The optimal solution to the correct QUBO.}\label{fig:undfeedvertex_correct}
	\end{subfigure}
\caption{Left: the undirected feedback vertex set in red, which is the optimal solution to the QUBO problem in \eqref{eq:undfeedvert} combined with \eqref{eq:symterm}, though an incorrect solution to the undirected feedback vertex set. Right: the undirected feedback vertex set in red, which is the optimal solution to the QUBO problem in \eqref{eq:undfeedvertcor} as well as to the undirected feedback vertex set.}
\end{figure}

A third issue is that the second term in \eqref{eq:undfeedvert} still is incorrect, even after the alteration described above. This term is set up such that, when $y_{uv}=y_{vu}=0$, the edge $(u,v)$ should be included in the `spanning forest' and thus it should have a depth. When either $u\in  F$ or $v\in F$ the edge $(u,v)$ should not have a depth by the third term in \eqref{eq:undfeedvert} and the extra term in \eqref{eq:symterm}. However, when both $u,v \in F$, that is, the edge $(u,v)$ points `towards' a vertex in the feedback set in both orientations $(u,v)$ and $(v,u)$, i.e.\ there is an edge connecting two vertices that should be in the feedback set, a problem arises. In this case, there is a penalty no matter the configuration of the sum $\sum_{i=1}^{\lfloor N/2\rfloor} (x_{uv,i}+x_{vu,i})$, since $1-(y_{uv}+y_{vu})$ then equals $-1$. An example where this goes wrong can be seen in Figure \ref{fig:undfeedvertex_wrong}, where the Hamiltonian used is the sum of the terms in \eqref{eq:undfeedvert} and \eqref{eq:symterm}. Here, the feedback set is shown in red, but the remaining graph still contains a cycle. This is exactly due to the issue described above, where it does not want to include vertex $2$ in the feedback set, as this will cause extra penalties as $y_{24}=y_{42}=y_{26}=y_{62}=1$ should hold.

In order to solve this issue, the term $1-(y_{uv}+y_{vu})$ should be replaced by something that equals $1$ if $y_{uv}=y_{vu}=0$, and equals $0$ otherwise. Thus, we replace it with $1-(y_{uv}+y_{vu}-y_{uv}y_{vu})$. That would mean the second term would become
\be 
A\sum_{(u,v)\in E}\Big(1-(y_{uv}+y_{vu}-y_{uv}y_{vu})-\sum_{i=1}^{\lfloor N/2\rfloor} (x_{uv,i}+x_{vu,i})\Big)^2.
\ee 
However, as products with more than two variables are not allowed, the square forms another issue. We can, however, simply omit it as configurations of the variables such that this term becomes negative can be sufficiently penalised elsewhere by altering the constant weights. Thus, the correct Hamiltonian should be of the form 
\be \begin{aligned}\label{eq:undfeedvertcor}
	H_A={}&A\sum_{v\in V} \Big(1-y_v-\sum_{i=0}^{\lfloor N/2\rfloor} x_{v,i}\Big)^2+A\sum_{(u,v) \in E} \Big((y_{uv}-y_{v})^2+(y_{vu}-y_{u})^2\Big)\\
	&+A\sum_{v\in V}\sum_{i=1}^{\lfloor N/2\rfloor}\Big(x_{v,i}-\sum_{u:(u,v)\in E}x_{uv,i}\Big)^2\\
	&+A\sum_{(u,v)\in E}\sum_{i=1}^{\lfloor N/2\rfloor} x_{uv,i}(2-x_{u,i-1}-x_{v,i})+x_{vu,i}(2-x_{v,i-1}-x_{u,i}),\\
	H_B={}&B\sum_{(u,v)\in E} \Big(\sum_{i=1}^{\lfloor N/2\rfloor} (x_{uv,i}+x_{vu,i})-(1-(y_{uv}+y_{vu}-y_{uv}y_{vu}))\Big),\\
	H_C={}&C\sum_{v\in V} y_{v},
\end{aligned}\ee 
where $A>B+2C,B>C>0$. Note that $H_B$ is negative only when $x_{uv,i}=x_{vu,i}=0$ for all $i\in\{1,\ldots,\lfloor N/2\rfloor\}$ and $y_{uv}=y_{vu}=0$ as well. This happens in one of two cases: $(1)$ all $x_{xu,i},x_{vu,i}$ equal $0$, but $y_{uv}=y_{vu}=0$ as well. This results in a penalty in the third term of $H_A$, or in the first or second if the $y_{v},y_{u},x_{v,i}$ and $x_{u,i}$ variables do not match the $y_{uv},y_{vu}$ variables, which counters the $-B$ value of $H_B$ and the decrease of $H_C$ by at most $2C$.

$(2)$ there is exactly on $x_{uv,i},x_{vu,i}$ that equals $1$, but at least one of the $y_{uv},y_{vu}$ equals $1$ as well. Again, this results in a penalty in one of the terms in $H_A$, which balances with the $-B$ value of $H_B$. It follows that the Hamiltonian does produce the correct optimal solution, as shown in Figure \ref{fig:undfeedvertex_correct}.

\subsubsection{Feedback edge set}

This problem is similar to the undirected feedback vertex set discussed in Section \ref{sec:undfeedvert}, though now the problems deals with a directed graph and the feedback set $F$ consists of edges that we want to delete, rather than vertices, to create an acyclic graph. \\

In a directed acyclic graph, it is possible to assign to each vertex a height, such that for any edge $(u,v)\in E$, the height of $v$ exceeds the height of $u$. We can construct the height of every vertex in a directed acyclic graph recursively. Namely, start with such a graph $G=(V,E)$ with $N$ vertices. Since it is acyclic, we can find a vertex $v$ with only outgoing edges (see \cite{Luc14} for the proof of this claim). Assign this vertex height $1$ (in case multiple vertices with only outgoing edges exist, choose any one of them). Then, repeat this step in the induced subgraph $G_1$, induced by $V\backslash\{v\}$, and assign the vertex you find here height $2$. Continue until this process only a single vertex is left, which you assign height $N$. Likewise, the edge-height starts from $1$ and an edge with edge-height $i$ connects a vertex of height $i$ to a vertex of greater height. 

\begin{figure}[h]
	\centering
	\begin{subfigure}{0.5\textwidth}
		\centering
		\includegraphics[width=0.6\textwidth]{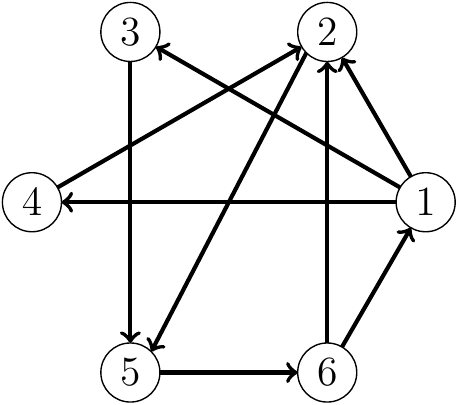}
		\caption{A sub-optimal solution to the incorrect QUBO in \eqref{eq:feededge} with lower $H$ value than the correct solution.}\label{fig:feededgeincorrect}
	\end{subfigure}%
	\begin{subfigure}{0.5\textwidth}
		\centering
		\includegraphics[width=0.6\textwidth]{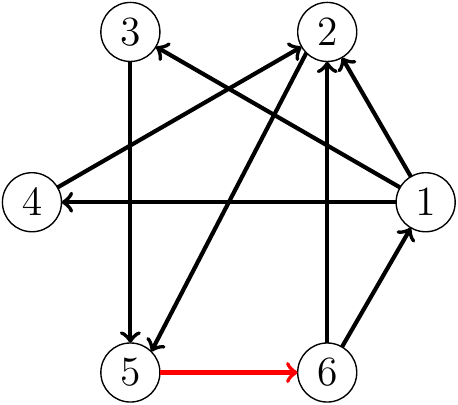}
		\caption{The optimal solution to the directed edge feedback set problem.}\label{fig:feededgecorrect}
	\end{subfigure}
	\caption{Left: a suboptimal solution to the incorrect QUBO in \eqref{eq:feededge} which has a lower Hamiltonian value than the optimal solution to the directed feedback edge set problem. Right: the optimal solution to both the correct QUBO in \eqref{eq:feededgecorrect} as well as the directed feedback edge set problem.}
\end{figure}

Lucas introduces the height variables $x_{v,i},v \in V,i \in \{1,\ldots,N\}$, where $x_{v,i}$ equals $1$ if vertex $v$ has height $i$. For edges, Lucas introduces the variables $y_{uv},(u,v)\in E$, where $y_{uv}$ equals $1$ if $(u,v)$ is \emph{not} in the feedback set, and $x_{uv,i},(u,v)\in E,i\in \{1,\ldots,N-1\}$, where $x_{uv,i}$ equals $1$ if $y_{uv}=1$, i.e.\ $(u,v)\not\in F$, and $(u,v)$ has edge-height $i$. Lucas then presents the Hamiltonian
\be \ba \label{eq:feededge}
H_A ={}& A\sum_v \Big(1-\sum_{i=1}^N x_{v,i}\Big)^2 + A\sum_{(u,v)\in E}\Big(y_{uv}-\sum_{i=1}^{N-1} x_{uv,i}\Big)^2 \\
&+ A\sum_{(u,v)\in E}\sum_{i=1}^{N-1} x_{uv,i}\Big(2-x_{u,i}-\sum_{j=i+1}^{N} x_{v,j}\Big),\\
H_B ={}& B\sum_{(u,v)\in E} (1-y_{uv}),
\ea \ee
with $A>B>0$. The mistake in this Hamiltonian is in the fact that the third term of $H_A$ can become negative, which yields incorrect optimal solutions. This term becomes negative if multiple $x_{v,j}$ equal $1$, which is not penalised heavily enough by the first term of $H_A$. An example where this Hamiltonian leads to incorrect solutions can be seen in Figure \ref{fig:feededgeincorrect}. The correct solution to the directed feedback edge set problem can be seen in Figure \ref{fig:feededgecorrect}. For this problem, using the variables described above, we can encode the optimal solution to the graph problem as follows. The variables set to equal $1$ are 
\be \ba \label{eq:varone}
x_{6,1},x_{1,2},x_{3,3},x_{2,4},x_{3,5},x_{5,6},\\
y_{12},y_{13},y_{14},y_{25},y_{35},y_{42},y_{61},y_{62},\\
x_{12,2},x_{13,2},x_{14,2},x_{25,4},x_{35,5},x_{42,3},x_{61,1},x_{62,1},
\ea\ee 
which corresponds to the vertices and edges having the correct height and edge-height, respectively, according to the procedure discussed above, and $F=\{(5,6)\}$. It is clear that this is indeed the optimal solution to the directed feedback edge set problem. However, it is not the optimal solution to the QUBO as described in \eqref{eq:feededge}, even though $H=H_A+H_B=B$ for these variables. Namely, let us set $x_{2,5}=1,y_{56}=1,x_{56,2}=1$. This would mean that $F=\emptyset$, as now the edge $(5,6)$ has $y_{56}=1$ and so it is no longer part of the feedback set. As we have assigned two different heights to vertex $2$, namely $4$ and $5$, we obtain a penalty of size $A$ from the first term of $H_A$. As both $y_{56}$ and one of the $x_{56,i}$ equal $1$, the second term of $H_A$ remains zero. Now, by inspecting the third term of $H_A$ for all edges ending in vertex $2$, we can see that this yields a decrease of $3A$, and thus $H_A=-2A$. Finally $H_B=0$, which is a decrease of $B$. So, overall, the Hamiltonian has decreased with $2A+B$, so $H=-2A$. We note that this is the optimal solution only if $A\geq 2B$; for $A\in(B,2B)$ the optimal solution would only set $x_{2,5}$ or $x_{2,6}$ to equal $1$ additionally to the variables in \eqref{eq:varone}, which yields a value of $H=-3A+B$. This would produce the correct feedback set, but the height assignment of the vertices would not be correct, which indicates that there might not be a choice of $A,B$ such that the correct optimal solution will be found for all graph instances.

A correct Hamiltonian would be
\be\ba \label{eq:feededgecorrect}
	H_A &= A\sum_{v\in V} \Big(1-\sum_{i=1}^N x_{v,i}\Big)^2+A\sum_{(u,v)\in E}\Big(y_{uv}-\sum_{i=1}^{N-1}x_{uv,i}\Big)^2,\\
	H_B &= B\sum_{uv\in E}\sum_{i=1}^{N-1}  x_{uv,i}\Big(2-x_{u,i}-\sum_{j=i+1}^{N}x_{v,j}\Big),\\
H_C &= C\sum_{uv\in E} (1-y_{uv}),
\ea\ee 
where $A>\Delta B,B>C>0$, where $\Delta$ is the maximum degree of the graph. We note that the configurations that yield a negative value of the Hamiltonian in \eqref{eq:feededge} yield a strictly positive value of the Hamiltonian in \eqref{eq:feededgecorrect}, whereas the optimal solution to the directed feedback edge set problem yields $H=0$ in both cases. \\

In general, using the new Hamiltonian, a $y_{uv}$ variable is not set to $1$ to decrease $H_C$ when this yields a penalty in $H_A$, and multiple $x_{uv,i},x_{v,i}$, for different values of $i$, are not set to $1$ either. Let us assume that we have a configuration such that $H_A=H_B=0$. We can then distinguish three cases in which $H_A$, $H_B$ change in value: $(1)$ for every edge $(u,v)$, at most one $x_{uv,i}$ equals $1$, but for some vertex $v$, multiple $x_{v,j}$ equal $1$, say $k$ many. This yields a decrease in $H_B$ of at most $(k-1)\Delta B$, but yields an increase in $H_A$ of $(k-1)^2A$. 

$(2)$ multiple $x_{uv,i}$ are set to equal $1$, say $\ell$ many, and, for some $v\in V$, $k$ many $x_{v,j}$ equal $1$. Then, $H_B$ decreases by at most $((k-1)+(\ell-1)(k-2))\Delta B$, $H_C$ decreases by at most $C$ and $H_A$ increases by $((\ell-1)^2+(k-1)^2)A$. 

$(3)$ multiple $x_{u,i}$ also equal $1$ on top of what is described in case $(2)$, say $\wt k$ many, in which case $H_B$ decreases by at most $(\wt k(k-1)+(\max(\ell-\wt k,0)(k-2))B$, $H_C$ decreases by at most $C$ and $H_A$ increases by $((\ell-1)^2+(k-1)^2+(\wt k-1)^d2)A$. All cases yield an overall increase in the Hamiltonian by the choice of the parameters, and thus the corrected Hamiltonian in \eqref{eq:feededgecorrect} maps the NP-hard problem to a QUBO correctly.\\

It could be possible to omit the variables $y_{uv}$ as well, that is, omit the second sum in $H_A$ and replace the $y_{uv}$ in $H_C$ by $\sum_{i=1}^{N-1}x_{uv,i}$, but that would require the constant weights to satisfy $A>N\Delta B,B>C>0$. Omitting the $y_{uv}$ requires less qubits to be used in QA devices but at the same time, can yield large ratios of constants in the terms of the Hamiltonian, which not ideal when embedding it into and using it for QA devices.

\section{QUBO formulations for additional NP-complete and NP-hard problems}\label{sec:map}

In this section we provide additional mappings from several NP-complete and NP-hard problems to QUBO problems. We do this in a similar form as the mappings discussed in Section \ref{sec:correction}, by providing a Hamiltonian which is to be minimised. Some of the problems are, as far as the author is aware of, not covered in other literature, and some generalise problems presented in, among others, \cite{Luc14}.
 
\subsection{Bin packing with integer weights}
In the bin packing problem, we are given $K$ objects with integer weights $w_j,j\in\{1,\ldots,K\}$ and $N$ bins, each with capacity $C\in\N$. The aim is to minimise the number of bins used, such that all objects are in a bin and no bin is filled beyond its capacity. Bin packing is a well-known NP-hard problem \cite{GarJohn90}.

In order to map this problem into a QUBO, we introduce the variables $x_{i,j},i\in\{1,\ldots,N\},j\in\{1,\ldots,K\}$, where $x_{i,j}=1$ if weight $j$ is placed in bin $i$, variables $x_i,i\in\{1,\ldots,N\}$, where $x_i=1$ if bin $i$ is not empty, and variables $y_{i,k},i\in\{1,\ldots,N\},k\in\{1,\ldots,C\}$, where $y_{i,k}=1$ if bin $i$ has been filled up to level $k$, that is, when the sum of the weights of the objects in bin $i$ equals exactly $k$. We can then formulate the Hamiltonian as
\be \ba 
H_A={}&A\sum_{i=1}^N\Big(x_i-\sum_{k=1}^C y_{i,k}\Big)^2+A\sum_{j=1}^K\Big(1-\sum_{i=1}^Nx_{i,j}\Big)^2+A\sum_{i=1}^N\Big(\sum_{k=1}^C k y_{i,k}-\sum_{j=1}^K w_j x_{i,j}\Big)^2\\
&+A\sum_{i=1}^N (1-x_i)\sum_{j=1}^K x_{i,j},\\
H_B={}&B\sum_{i=1}^N x_i,
\ea \ee  
with $A>2B>0$. The first term of $H_A$ ensures that every bin that is used is filled up to a unique level, and every unused bin is not filled to any level at all. By the second term, every object needs to be allocated to a bin. The third term penalises configurations where bins are filled beyond their capacity and the last term of $H_A$ ensures only non-empty bins are counted as used. Finally, $H_B$ represents the number of bins used. \\

The requirement $A>2B$ is to ensure that one never fills bins beyond their capacity in favour of using fewer bins. Given a capacity $C$, the minimal penalty to `increase' the capacity the most is $A$ and the capacity can then be increased to $C+C-1=2C-1$. This is realised by setting $y_{i,C}=y_{i,C-1}=1$, so that the first term of $H_A$ yields a penalty of $A$. Let us now consider the following example: for $C,N\in\N$, we have $3N$ bins and $K=3N$ objects, each of weight $\lceil (C+1)/2 \rceil$, such that every object needs to be placed into a unique bin. As above, we set $y_{i,C}=y_{i,C-1}=1$ for $i=1 \ (\!\!\!\!\mod 3)$ and all other $y_{i,k}$ are set to zero. As we can now fill every third bin beyond capacity, the number of bins required will decrease. Since all objects have weight $\lceil (C+1)/2 \rceil$ and the capacity of every third bin is now $2C-1$, it follows that we can place at most $3$ objects in these bins, since $(2C-1)/\lceil (C+1)/2 \rceil<4$. Thus, the number of required bins has decreased from $3N$ to $N$. Since we can set the variables $x_{i,j},x_i$ in the proper way so that no extra penalty is incurred in $H_A$, we end up with $H_A=NA, H_B=NB$, so that the total value of the Hamiltonian equals $N(A+B)$. In order for this solution to be sub-optimal, it follows that $N(A+B)>3NB$ is required, or $A>2B$.\\ 

Finally, it is also not favourable to not allocate weights to a bin, as this results in a penalty of at least $A$ times the number objects that are not allocated, which is larger than $2B$ times the number of bins that end up being unused as a result.

\subsection{Partitioning problems}
\subsubsection{Number partitioning}
Number partitioning deals with splitting a set $S=\{s_1,s_2,\ldots,s_N\}$ of $N$ positive integers into $m$ disjoint subsets $S_j,j\in\{1,\ldots,m\}$, such that the sum of the integers in each set is the same. This is a generalisation of the number partitioning problem as described in \cite{Luc14}, where only the case $m=2$ is dealt with. For the general case, we can introduce variables $x_{i,j}$, where $x_{i,j}$ equals $1$ if $s_i\in S_j$. Then, we can construct the Hamiltonian
\be \ba
H_A&=A\sum_{i=1}^N \Big(1-\sum_{j=1}^m x_{i,j}\Big)^2\\
H_B&=B\sum_{1\leq j_1<j_2\leq m} \Big(\sum_{i=1}^N s_i x_{i,j_1} -\sum_{i=1}^N s_i x_{i,j_2} \Big)^2,
\ea\ee 
where $A>Bm\max_{1\leq i\leq N}s_i^2,B>0$. The first term ensures that every number is part of exactly one subset and thus the subsets $S_j$ are disjoint, and the second term penalises configurations where the numbers in the different subsets do not have the same sum. It is clear that $H=0$ if and only if the sum of the integers in each subset is equal and every number belongs to exactly one subset.

\subsubsection{Graph partitioning}
In a similar way to the number partitioning problem, we can generalise graph partitioning into $m\geq 2$ sets, as described when $m=2$ in \cite{Luc14}, as well. For a graph $G=(V,E)$, we now aim to split the vertices into $m$ disjoint subsets $V_j,j\in\{1,\ldots,m\}$, such that the subsets are of equal size and the number of edges between the subsets is minimised. We set the variable $x_{v,j}$ equal to $1$ if $v\in V_j$, $v\in V,j\in\{1,\ldots,m\}$. We obtain a similar Hamiltonian of the form
\be \ba
H_A&=A\sum_{v\in V} \Big(1-\sum_{j=1}^m x_{v,j}\Big)^2,\\
H_B&=B\sum_{1\leq j_1<j_2\leq m} \Big(\sum_{v\in V} x_{v,j_1} -\sum_{v\in V} x_{v,j_2} \Big)^2,\\
H_C&=C\sum_{j=1}^m \sum_{(u,v)\in E} x_{u,j}+x_{v,j}-2x_{u,j}x_{v,j},
\ea \ee  
where $ A>B> C\min(m\Delta,N)/(m(m+2)),C>0$. $H_A$ penalises configurations where a vertex belongs to multiple subsets, $H_B$ penalises configurations where the subsets are not of equal size, and $H_C$ penalises configurations in which many edges between the subsets are present. By the choice of the parameters, a change in the variable values, leading to a decrease in $H_C$ and/or $H_B$ is always met with an increase in $H_A$ and/or $H_B$, so that from the optimal value of $H=H_A+H_B+H_C$ we obtain the optimal graph partitioning. The choice of the parameters follows from an argument analogous as the argument made in \cite{Luc14}, where shifting a vertex from one subset to another to decrease $H_C$ yields a decrease of at most $\min(\Delta,N/m)C$, whilst $H_B$ increases at least $4B+(m-2)B=(m+2)B$.

\subsubsection{Subset sum problem}
Related to number partitioning is the subset sum problem. In this problem, we are given a set $S=\{s_1,s_2,\ldots,s_N\}$ of $N$ integers and a target value $t\in \Z$, and the aim is to find a subset of $S$ such that the elements of this subset sum to $t$. The subset sum problem is related to the knapsack problem \cite{Dantz57} and is one of Karp's NP-complete problems \cite{Karp72}. For this problem we introduce variables $x_i,i\in\{1,\ldots,N\}$, where $x_i$ equals $1$ if $s_i$ is included in the subset. A Hamiltonian of the form
\be 
H=\Big(\sum_{i=1}^N s_ix_i-t\Big)^2
\ee 
yields the optimal value zero if and only if the sum of the numbers included equals $t$.

\subsection{Numerical three-dimensional matching}
In the numerical three-dimensional matching problem, we are given three equally sized sets $X:=\{x_1,x_2,\ldots,x_N\},Y:=\{y_1,y_2,\ldots,y_N\},Z:=\{z_1,z_2,\ldots,z_N\}$ of integers and an integer $b$, and the task is to find a set $M\subset X\times Y\times Z$, such that every element of $X,Y,Z$ occurs exactly once in $M$, and such that for all $m=(m_1,m_2,m_3)\in M$, $m_1+m_2+m_3=b$. The problem is NP-complete, and is related to the $3$-dimensional matching problem \cite{GarJohn90}.

In order to map it to a QUBO, we introduce the variables $x_{i,j},y_{i,j},z_{i,j},i,j\in\{1,\ldots,N\}$, where $x_{i,j},y_{i,j},z_{i,j}$ equals $1$ if $x_i\in X,y_i\in Y,z_i\in Z$ is the first, second and third element of the $j^{\text{th}}$ element of $M$, respectively. We then construct the Hamiltonian
\be \ba
H_A={}&A\sum_{i=1}^N\bigg[\Big(\sum_{j=1}^N x_{i,j}-1\Big)^2+\Big(\sum_{j=1}^N y_{i,j}-1\Big)^2+\Big(\sum_{j=1}^N z_{i,j}-1\Big)^2\bigg]\\
&+A\sum_{j=1}^N\bigg[ \Big(\sum_{i=1}^N x_{i,j}-1\Big)^2+\Big(\sum_{i=1}^N y_{i,j}-1\Big)^2+\Big(\sum_{i=1}^N z_{i,j}-1\Big)^2\bigg],\\
H_B={}&B\sum_{j=1}^N\Big(\sum_{i=1}^N (x_ix_{i,j}+y_iy_{i,j}+z_iz_{i,j})-b\Big)^2,
\ea\ee 
where $A>B\sum_{S\in\{X,Y,Z\}}\max_{s\in S}s^2, B>0$. The first double sum in $H_A$ ensures that every element in $X,Y,Z$ is found in exactly one element of $M$, the second double sum ensures that to every element in $M$ exactly one element of $X,Y,Z$ is assigned. $H_B$ then penalises configurations such that the elements in $M$ do not sum to $b$.

\subsection*{Acknowledgements}
The research presented in this paper has been conducted at the Applied Research department of BT in Ipswich, England. The author would like to thank Keith Briggs for many helpful discussions and pointing out the paper discussed here.

\bibliographystyle{abbrv}
\bibliography{QUBOref}	
\end{document}